# Narrow Inhomogeneous Distribution and Charge State Stabilization of Lead-Vacancy Centers in Diamond

Ryotaro Abe[1], Peng Wang[1], Takashi Taniguchi[2], Masashi Miyakawa[2], Shinobu Onoda[3], Mutsuko Hatano[1], Takayuki Iwasaki[1,*]

[1]Department of Electrical and Electronic Engineering, School of Engineering, Institute of Science Tokyo, Meguro, 152-8552 Tokyo, Japan

[2]Research Center for Materials Nanoarchitectonics, National Institute for Materials Science, 305-0044 Tsukuba, Japan

[3]Takasaki Advanced Radiation Research Institute, National Institutes for Quantum Science and Technology, 1233 Watanuki, Takasaki, 370-1292 Gunma, Japan

[*]Email: iwasaki.t.c5b4@m.isct.ac.jp

**Abstract**

Lead-vacancy (PbV) centers in diamond with a large ground state splitting are expected to be a building block of quantum network nodes. Due to the heaviness of the Pb atom, it is challenging to fabricate high-quality PbV centers with a narrow inhomogeneous distribution and stable charge state. In this study, for the formation of the PbV centers, high temperature anneal up to 2300°C is performed after Pb ion implantation. At a lower temperature of 1800°C, the PbV centers show a large inhomogeneous distribution and spectral diffusion, while higher temperatures of 2200-2300°C leads to narrow inhomogeneous distributions with standard deviations of ~5 GHz. The charge state transition of the PbV centers formed at 2200°C occurs by capturing photo-carriers generated from surrounding defects under 532 nm laser irradiation. Finally, multiple stable PbV centers with nearly identical photon frequencies are obtained, which is essential for applications in quantum information processing.

# 1. Introduction

Color centers in diamond have attracted significant attention to distribute quantum entanglement between distant quantum network nodes. Entanglement generation between remote nodes has been reported using nitrogen-vacancy (NV) centers[1,2] and silicon-vacancy (SiV) centers[3]. However, the NV centers face challenges such as a low concentration of the fluorescence into the zero-phonon line (ZPL) and spectral instability. In contrast, group-IV vacancy centers exhibit high ZPL concentration and spectral stability due to their inversion symmetry[4–6]. The SiV center has a low quantum efficiency[7] and the spin coherence is limited by phonon-induced transitions between ground states, requiring SiV and GeV centers cooling to mK temperatures to achieve coherence times in the millisecond range[8,9]. A lead-vacancy (PbV) center, another kind of group-IV vacancy center, is expected to achieve a millisecond-scale spin coherence time at 9 K owing to the large ground-state splitting[10], which does not require a dilution refrigerator. The ZPLs of the PbV center appear at around 550 nm with the capability of single photon emission[10,11]. Recently, a transform-limited linewidth under resonant excitation has been also demonstrated[12]. However, no PbV centers with an identical fluorescence wavelength (photon frequency) under resonant excitation have been reported. Due to the heaviness of the Pb atoms, it is challenging to fabricate PbV centers with a narrow inhomogeneous distribution and an identical photon frequency under resonant excitation, which is essential for protocols in quantum information processing[13–15].

Imperfection of the diamond lattice, i.e. strain and defects, hampers the fabrication of stable quantum emitters with an identical photon frequency. Figure 1 illustrates several conditions of quantum emitters in a solid-state material. Strain in the host material modulates the energy level of each emitter, leading to non-uniform distinguishable photon frequencies (Figure. 1(a)). Additionally, when a defect in vicinity, referred to X, has dynamic change in its charge state under laser irradiation, the generated local electric-field leads to the Stark shift of nearby group-IV vacancy centers[16], causing spectral diffusion (Figure 1(b)). Beyond the energy shift, a defect in the lattice, referred to Y, possibly generate photo-carriers under laser irradiation, and then, emitters capturing the carriers transition to another charge state[17], as shown in Figure 1(c). Therefore, the formation of multiple quantum emitters with an identical photon frequency under resonant excitation (Figure 1(d)) requires both suppressing the inhomogeneous distribution and ensuring the charge stability of the quantum emitters.

In this study, we investigate the inhomogeneous distribution and charge stability of PbV centers formed by high-temperature annealing at 1800-2300°C. Our findings indicate that an annealing temperature of 1800°C is insufficient to achieve an inhomogeneous distribution comparable to that of other group-IV vacancy centers, whereas temperatures in the range of 2200–2300°C are adequate to obtain narrow inhomogeneous distributions. Furthermore, in a sample annealed at a temperature of 2200°C, we observe the dark state transition of the PbV centers due to the capturing of photo-carriers generated under non-resonant green laser irradiation. This phenomenon does not occur in a 2300°C annealed sample, suggesting that the number of residual defects is sensitive to the annealing temperature. Finally, we observe multiple PbV centers with identical resonant photon frequencies and transform-limited linewidths.

## 2. Results and discussion
### 2.1. Narrow inhomogeneous distribution of PbV centers with transform-limited linewidth

A PbV center consists of a Pb atom positioned between two adjacent vacancies,[6] as illustrated in Figure 1(e). The negatively charged PbV (PbV$^-$) center forms an S=1/2 system, where both the ground and excited states are split by the spin-orbit interaction, resulting in the energy level diagram shown in the inset of Figure 1(f). Accordingly, there are four optical transitions, labeled A to D. However, the A- and B-transitions are not observed because thermal excitation to the upper branch of the excited state is insufficient, and thus only the C- and D-transitions are observed[10] (Figure 2(b)). Additionally, the linewidth of the D-transition becomes four orders of magnitude broader than the C-transition at a low temperature due to the phonon-induced relaxation in the ground state[12]. Therefore, we focus on the C-transition in this study. Figure 1(f) shows a fluorescence decay of the PbV centers formed at 2300°C. From fitting with a single exponential function, we obtain a lifetime of $\tau$ = 4.4 ns, corresponding to the transform-limited linewidth of 36.2 MHz, which agrees with a previously reported linewidth[12]. We confirm that the PbV center annealed at 2200°C have a similar excited state lifetime.

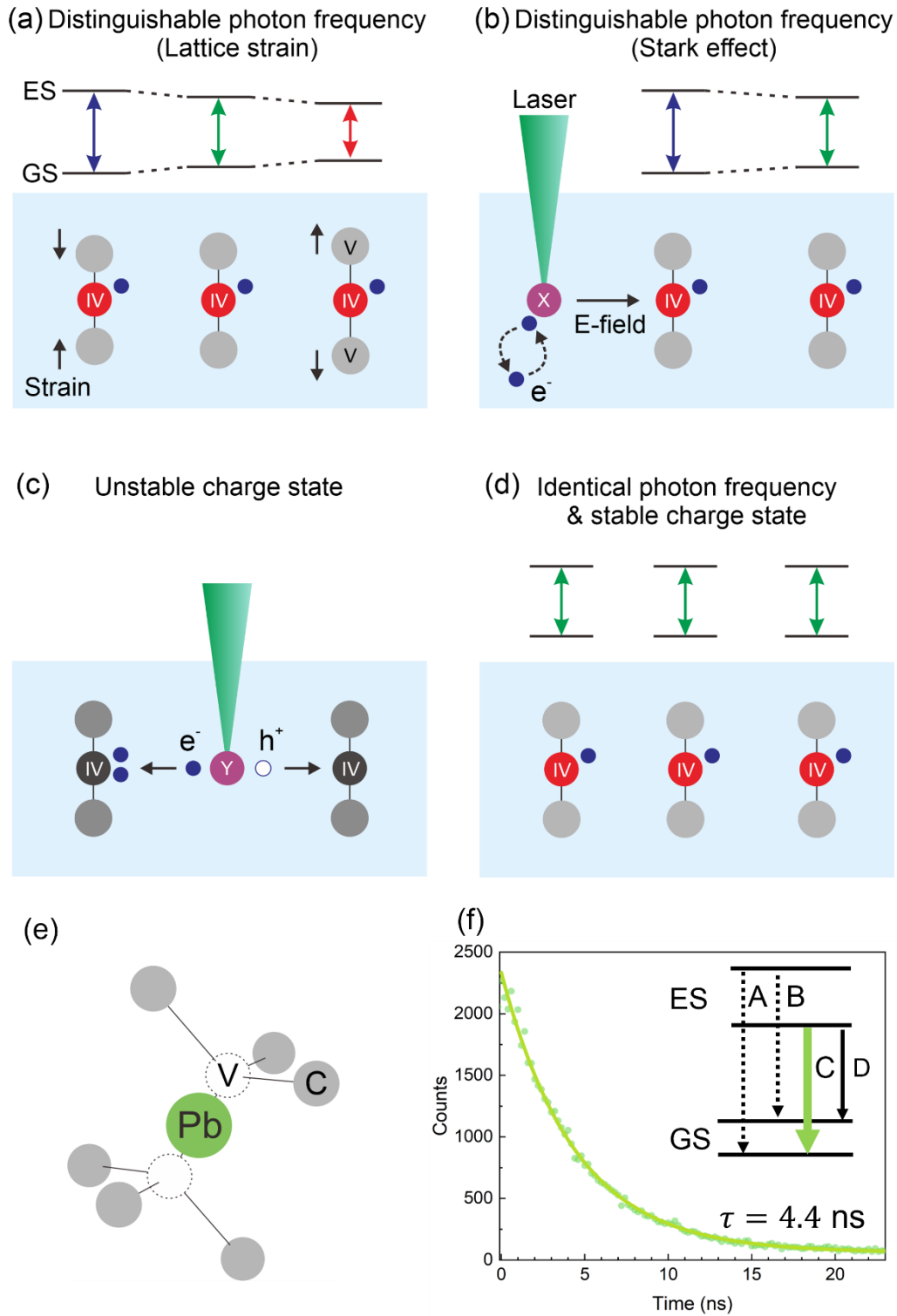

**Figure 1.** Quantum emitters in a solid-state material and properties of the PbV center. (a) Distinguishable frequencies from each emitter due to the lattice strain. The emitters with modified energy levels produce the corresponding photon frequencies, mentioned as different color arrows. GS and ES levels denote ground and excited states, respectively. (b) Distinguishable frequencies due to the Stark effect. Laser irradiation dynamically changes the charge state of a nearby defect (X) and an electric-field is locally generated to modulate the energy level and cause spectral diffusion. (c) Charge state transition by capturing an electron or a hole from an optically-excited defect (Y). The photo-carriers generated from the defects under laser irradiation diffuse to emitters. When -1 negatively charged emitters capture photo-carriers, they turn to another charge state, e.g. neutral or -2 negatively charged state. (d) Multiple emitters with an identical photon frequency and stable charge state. (e) Atomic structure of a PbV center with the main axis into the <111> direction. (f) Fluorescence decay of the PbV center. Note that we do not confirm whether the measurement spot is a single PbV center or not. Inset: energy levels of the PbV center.

Figure 2(a), (d), and (g) show photoluminescence excitation (PLE) spectra of PbV centers formed at different anneal temperatures of 1800°C, 2200°C, and 2300°C, respectively, under high-pressure. Hereinafter, each sample is referred as Sample 1800, Sample 2200, and Sample 2300. In Figure 2(a), solid lines represent each scan of three consecutive PLE measurements and a dashed line represents the averaged spectrum of the three scans. The linewidth in each scan is approximately 40 MHz, which is close to the transform-limited linewidth of the PbV center, while a spectral diffusion of 30-60 MHz occurs in every scan, resulting in a broadened linewidth of 93 MHz in the averaged spectrum. This spectral diffusion is thought to be caused by the Stark effect under the laser irradiation (Figure 1(b)). It is worth noting that, only these spectra are observed in PLE trials for this PbV center, and subsequent attempts to detect a PLE spectrum at the same wavelength after green initialization laser is unsuccessful. Also, we do not observe any PLE spectrum for other PbV centers in this sample. These facts indicate that the PbV centers formed at 1800°C are unstable with respect to the surrounding environment and their own charge state. Figure 2(d) and (g) show the averaged PLE spectra of three and four scans for emitters in Sample 2200 and Sample 2300, respectively, showing narrow linewidths of approximately 39 - 40 MHz. Despite averaging multiple scans, the linewidth remains close to the transform-limited linewidth, indicating suppressed spectral diffusion in these PbV centers. Distributions of the PLE linewidth of Sample 2200 and Sample 2300 in Figure 2(e) and (h) statistically confirm the narrow linewidth of these two samples. Note that data of multiple scans that average to a single peak are evaluated by fitting, while data with large spectral diffusion that show multiple peaks after averaging are excluded from the histograms (Supporting Information, Figure S1). The number of the excluded data is 5 out of 77 PbV centers in Sample 2200 and 1 out of 94 PbV centers in Sample 2300. This means that although broad linewidths over 60 MHz are present in Sample 2300, even larger spectral diffusion is likely to occur in Sample 2200.

Then, we examine the inhomogeneous distribution of each sample. Figure 2(b) shows PL spectra at two laser spots in Sample 1800. One spot shows a typical spectrum with a ground state splitting of

~3900 GHz, while on another spot, the splitting increases to ~6000 GHz and both the C and D peaks largely shift, caused by the lattice strain around the emitter (Figure 1(a)). The inhomogeneous distribution of the C-peak photon frequency has a large standard deviation of 1550 GHz (Figure 2(c)). Then, due to the limited resolution of the spectrometer of the PL measurement, the inhomogeneous distributions of Sample 2200 and Sample 2300 are derived from the PLE measurements. As shown in Figure 2(f) and (i), we obtain much narrower standard deviations of 5.4 GHz and 4.9 GHz, indicating that the PbV centers are formed in a more uniform strain environment. The probability that two PbV centers are found within the transform-limited linewidth is evaluated to be 0.43% for Sample 2200 and 0.65% for Sample 2300. The inhomogeneous distributions obtained here are comparable to that of other group-IV vacancy centers formed by ion implantation and annealing[18–20]. A PbV sample, fabricated at a lower acceleration energy and a higher fluence for Pb ion implantation, has been shown to possess a much larger standard deviation of 66 GHz in the inhomogeneous distribution[10]. Therefore, the narrow inhomogeneous distribution obtained here suggests that the high energy ion implantation and the high-pressure and high-temperature (HPHT) annealing can produce high-quality PbV centers even for heavy Pb atoms, by suppressing the strain (Figure 1(a)) and Stark effect from surrounding defects (Figure 1(b)). It is worth mentioning that although $^{208}$Pb ions are mainly implanted in the samples, there is a possibility that isotopes with similar masses are mixed into the samples. Due to the heaviness of the Pb atom, the photon frequency shift by the Pb isotope becomes smaller compared to SiV, GeV, and SnV centers[18,21–23]. Small peaks at higher frequencies in Figure 2(f, i) might correspond to the isotope shift, suggesting that a single isotope has a narrower distribution.

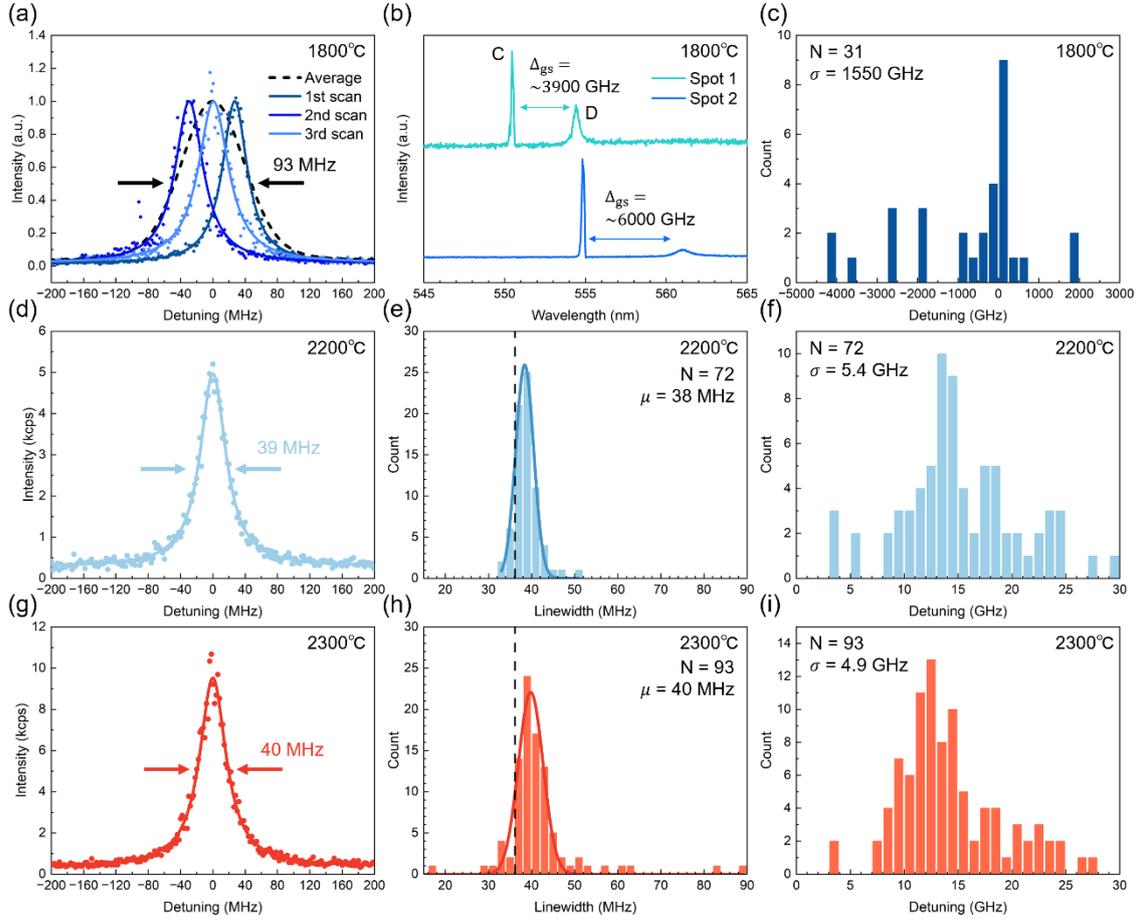

**Figure 2.** Linewidth and inhomogeneous distributions. (a) PLE spectra of a PbV center in Sample 1800. Three consecutive single scans and the averaged spectrum are displayed. The linewidths of each single scan are 37 MHz (1st), 42 MHz (2nd), and 45 MHz (3rd), while the linewidth of the averaged spectrum is broadened to be 93 MHz due to spectral diffusion. The fitting is performed using a Lorentzian function for each scan (solid line) and a Voigt function for the averaged spectrum (dashed line). The zero detuning corresponds to the center of the averaged spectrum. The spectra are normalized to the maximum of each fitting. (b) PL spectra at two different spots in Sample 1800. (c) Inhomogeneous distribution in Sample 1800, obtained from the PL measurements. (d) Averaged PLE spectrum of three scans of a PbV center in Sample 2200. (e) Linewidth distribution in Sample 2200. (f) Inhomogeneous distribution in Sample 2200. (g) Averaged PLE spectrum of four scans of a PbV center in Sample 2300. (h) Linewidth distribution in Sample 2300. (i) Inhomogeneous distribution in Sample 2300. Dashed lines in panels (e, h) denote a transform-limited linewidth of 36.2 MHz.

**2.2. Charge state transition by capturing photo-carriers**

Owing to the narrow inhomogeneous distribution, we find two PbV centers with close photon frequencies in Sample 2200 (Figure 3(a)), mentioned as PbV 1 and PbV 2 in the confocal fluorescence microscopy (CFM) image under non-resonant 532 nm laser in Figure 3(b). Since their PLE spectra overlap, both PbV centers are expected to be excited at the resonance frequency at the zero detuning (dashed line). However, only one PbV center (PbV 1) is observed in the image obtained in a resonant laser scan at the zero-detuning frequency after focusing on PbV 1 by 532 nm initialization laser (Figure 3(c)). A bright spot corresponding to PbV 1 is clearly observed, but no emitter is detected at the position of PbV 2. A similar experiment is conducted by focusing the non-resonant laser on PbV 2 and then scanning with the resonant laser at the same frequency (Figure 3(d)). In this case, PbV 2 is observed while the spot of PbV 1 becomes dark. Finally, when the non-resonant laser is focused on a different location, away from the two PbV centers and then resonant scan is performed, neither PbV 1 nor PbV 2 is observed (Figure 3(e)). These observations suggest that in this sample, the PbV centers changes to another dark state when the 532 nm non-resonant laser is irradiated to a region other than the target PbV center, and thus the emitter loses the bright fluorescence upon original resonant excitation. It is thought that the initialization of the target PbV to the bright negatively charged state is more effective compared with the fluorescence termination when 532 nm laser is directly irradiated to the target emitter, enabling us to observe only one PbV center in Figure 3(c, d).

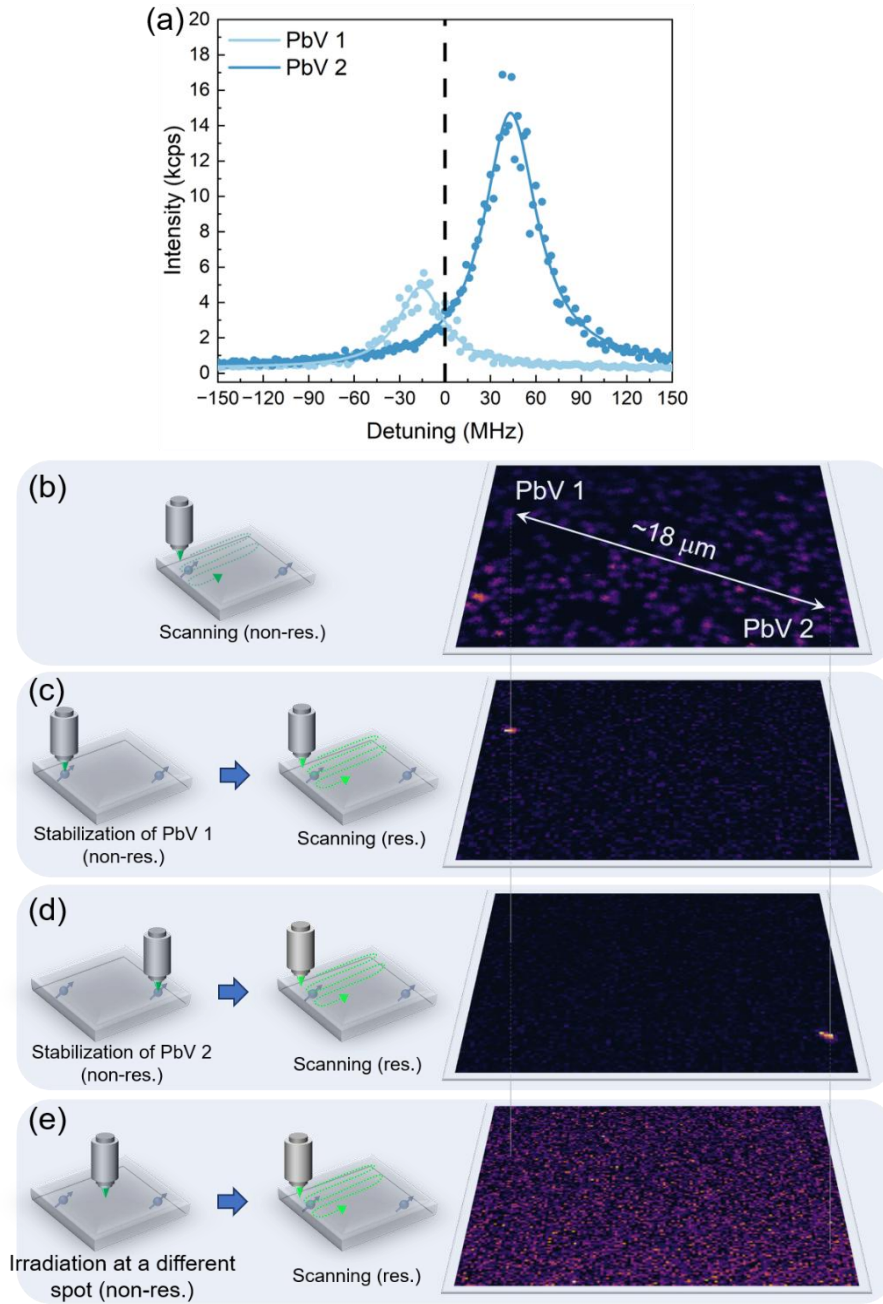

**Figure 3.** Instability of bright charge state in Sample 2200. (a) PLE spectra of two PbV centers (PbV 1, PbV 2 in panel (b)). (b) 532 nm non-resonant CFM image containing the two PbV centers, ~18 μm apart. (c) Resonant CFM image, after irradiating PbV 1 with 532 nm non-resonant laser. (d) Resonant CFM image, after irradiating PbV 2 with 532 nm non-resonant laser. (e) Resonant CFM image, after irradiating a point other than PbV 1 and PbV 2 with 532 nm non-resonant laser. The resonant scan images in panels (c-e) are obtained at the zero-detuning frequency of 544.5510 THz in panel (a).

To further investigate this phenomenon in detail, we conduct experiments by controlling the non-resonant laser irradiation. As shown in Figure 4(a), after stabilizing the charge state of a target PbV center using the 532 nm non-resonant laser, we irradiate the non-resonant laser at a spot approximately 4.5 μm away from the target PbV center, where neither NV nor PbV center are present (Supporting Information, Figure S2). The non-resonant laser is turned off after the irradiation, then we move to each position. Finally, we conduct three consecutive PLE scans of the target PbV center using the resonant laser. This process is repeated ten times at each non-resonant laser irradiation energy, defined as the product of the laser power and irradiation time. Figure 4(b) and (c) show the ten PLE spectra when the non-resonant laser energy is 0.375 mJ for Sample 2200 and 1.5 mJ for Sample 2300, respectively. In Sample 2200, a resonant peak is observed in only 4 out of 10 sequences, whereas in Sample 2300, it is observed in 8 out of 10 sequences. Since the PLE scans are conducted in a range of approximately ±1 GHz around the center frequency, spectral diffusion is unlikely to cause the undetectable peak. Figure 4(d) shows the probability of observing the resonant peak as a function of the non-resonant laser energy applied to the spot away from target PbV center (middle of Figure 4(a)). In Sample 2200, the probability decreases as increasing the irradiation energy. At 0.75 mJ or higher, the PLE spectrum is no longer observed at all. In contrast, in Sample 2300, the probability remains stable with the average value of 84%, even when the irradiation energy varies by an order of magnitude. The disappearance of PLE spectra in Sample 2300 is thought to correspond to the population of the negatively-charged state of the PbV center upon non-resonant 532 nm laser excitation[24], not mainly originating from the 532 nm laser irradiation at a spot away from the PbV center.

The more frequent disappearance of the PLE peak in Sample 2200 can be attributed to photo-carriers generated from defects in the sample (Figure 1(c)). When excited by the 532 nm non-resonant laser, some kinds of defects should produce photo-carriers which diffuse through the diamond lattice and are eventually captured by PbV centers, altering their charge state, as observed in NV and SiV centers[17,25–28]. If a negatively-charged PbV center captures an electron (hole), it changes to $PbV^{2-}$ ($PbV^0$). The number of generated carriers should increase with a higher laser energy until saturation, allowing them to affect more PbV centers. Since this phenomenon occurs when irradiating an area that does not have NV and PbV centers, photo-carriers here are thought to be generated from other defects. Alternative candidate would be substitutional nitrogen atoms (P1 center), Pb-related defects except PbV center, and vacancy clusters. Further studies are necessary to clarify defects generating photo-carriers. We note that another sample annealed at 2200°C show the opposite result as the sample above, where the PbV center is stable under the 532 nm laser irradiation at a spot away from a target PbV center (Supporting Information, Figure S3), fabricated even under the same sample preparation conditions as those for the unstable sample. These results suggest that the threshold temperature for the annealing effect on defect formation suppression may lie between 2200°C and 2300°C. At high-pressure annealing temperatures above 2000°C, temperature gradients within the sample cell and variations in the pressure generated during each experiment may cause annealing temperature errors of approximately ±30°C. To gain a deeper understanding of phenomena sensitive to annealing temperature, such as the one observed in this study, we need more precise evaluation of the generated temperature distribution in the high-pressure cell and the reproducibility (also see Experimental Section).

We summarize the distribution of the central frequency and linewidth of the PLE spectra obtained for Sample 2200 (Figure 4(e, f)) and Sample 2300 (Figure 4(g, h)) after irradiation at each irradiation energy. In both samples, the center frequencies are distributed around the zero detuning, and the linewidths remain close to the transform-limited linewidth, indicating that the PbV centers are stable from the point of view of spectral diffusion, even though the generation of photo-carriers clearly occurs in Sample 2200.

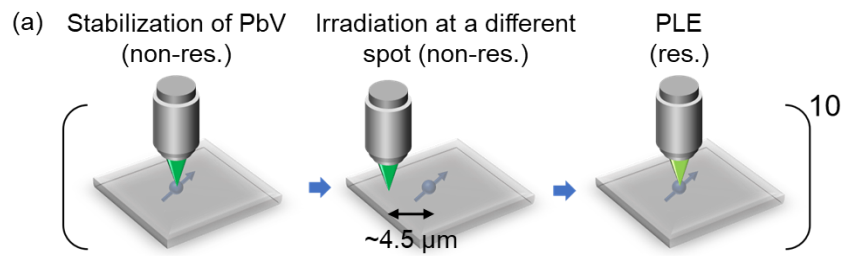
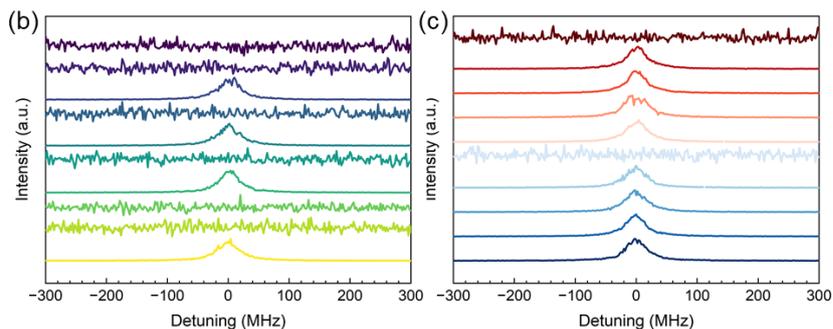
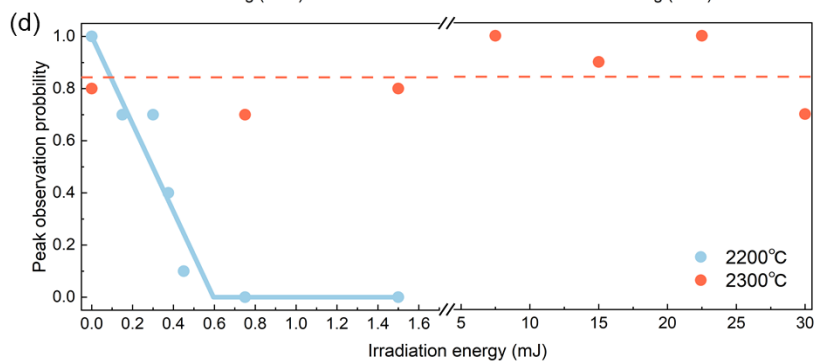
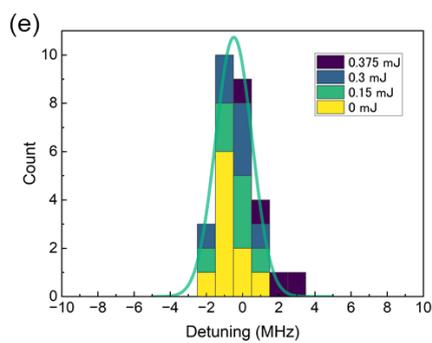
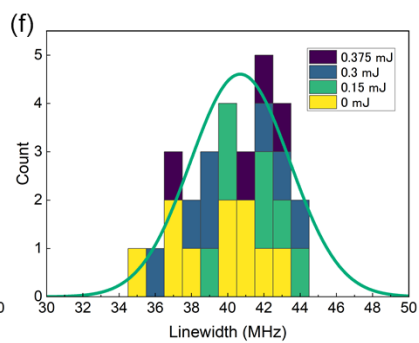
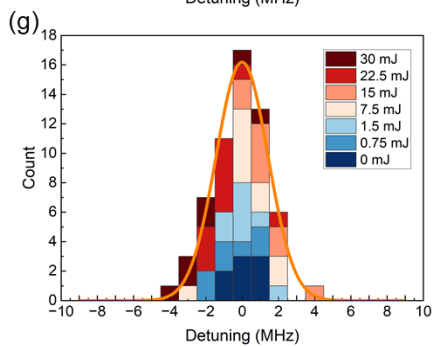
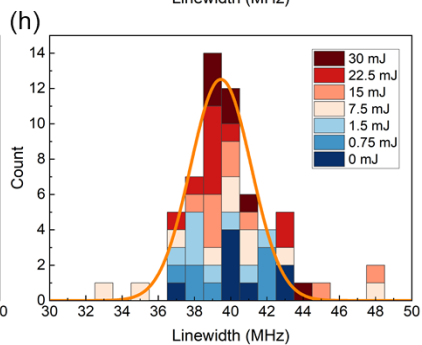

**Figure 4.** Photo-carrier generation and stability of PbV centers. (a) Measurement protocol with a controlled 532 nm non-resonant irradiation. After stabilizing a target PbV center with the non-resonant laser (left), the objective lens is moved to a location where there are no NV or PbV centers, and the controlled non-resonant laser is irradiated (middle). Finally, the objective lens is returned to the target PbV center, and three consecutive PLE scans are performed (right). This sequence is repeated ten times for each non-resonant laser energy. (b) Ten PLE spectra in Sample 2200 at a non-resonant energy of 0.375 mJ. (c) Ten PLE spectra in Sample 2300 at a non-resonant energy of 1.5 mJ. (d) Probability of observing the resonant peak as a function of the irradiation energy of the 532 nm laser. Red dashed line represents the average probability for Sample 2300. We change the laser power in the range of 0-1.5 mJ at a fixed time of 5 s, while the irradiation time is varied for the higher energy at a fixed power of 300 μW. Distributions of (e) central frequency and (f) linewidth of PbV centers in Sample 2200. Distributions of (g) central frequency and (h) linewidth of PbV centers in Sample 2300. The zero detuning corresponds to the central frequency of the first observed PLE spectrum at each energy. There is only one date point for 0.45 mJ in Sample 2200, so it is not included in the histograms in panels (e, f).

## 2.3. Multiple PbV centers with nearly identical photon frequencies

Finally, we demonstrate multiple PbV centers with nearly identical photon frequencies owing to the narrow inhomogeneous distribution. Figure 5(a) shows a CFM image of Sample 2300 obtained with resonant scanning at a specific frequency, where several bright spots are simultaneously observed, in contrast with the observations in Figure 3. Six PbV centers marked in circle show PLE spectra with similar center photon frequencies and narrow linewidths of 38 - 43 MHz (Figure 5(c)). Notably, two spectra have a very small energy difference of only 1 MHz. These PbV centers are expected to exhibit a high visibility in two-photon interference[13]. Another sample fabricated 2300°C also show multiple PbV centers with overlapped PLE spectra (Figure 5(b, d)). It is worth noting that multiple emitters are also resonantly observed in Sample 2200 when a weak 532 nm laser power is utilized before the resonant scan (Supporting Information, Figure S4). The charge stable and indistinguishable multiple quantum emitters are an important platform for the generation of graph states[29–33]. Thus, the PbV material developed in this study will become a new platform for quantum information processing.

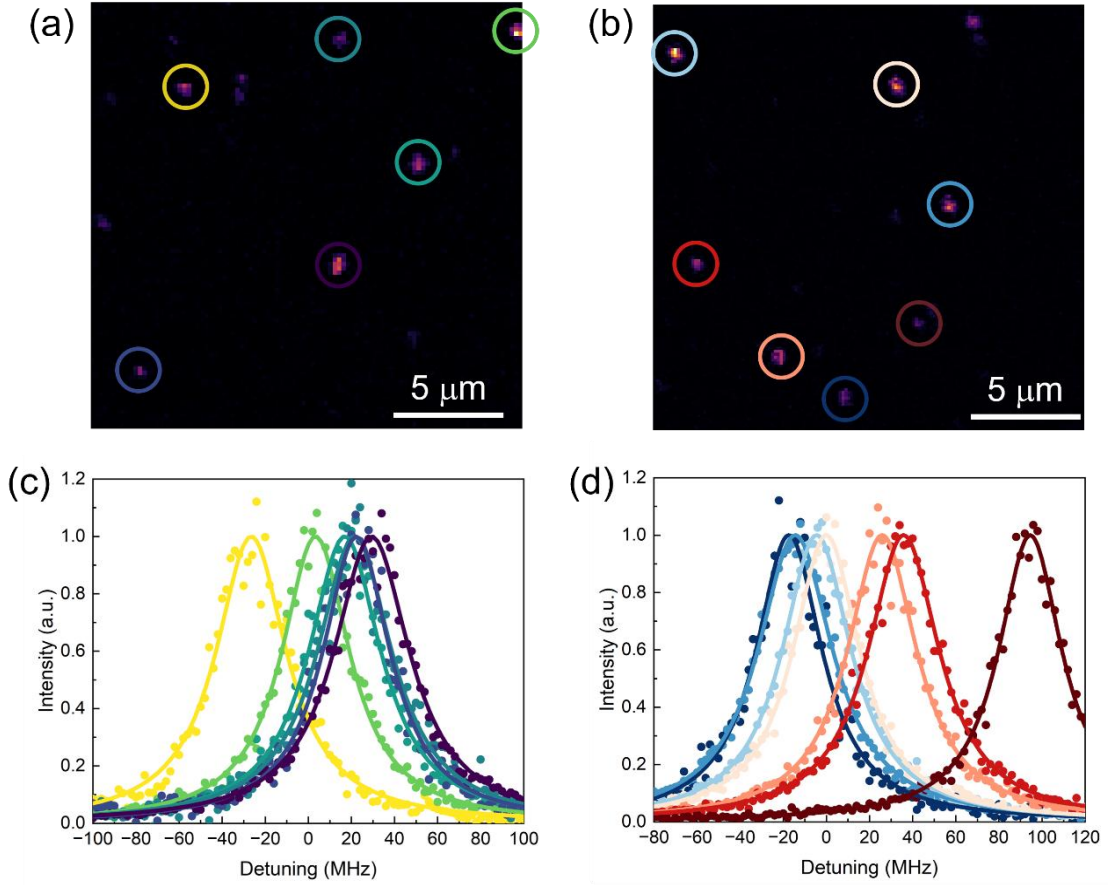

**Figure 5.** Multiple PbV centers with nearly identical photon frequencies and transform-limited linewidths. (a, b) CFM images obtained by resonant excitation at certain frequencies for two different samples annealed at 2300°C. (c, d) PLE spectra from bright spots in panels (a, b). The spectra are normalized to the maximum of each fitting. The zero-detunings are (c) 544.5495 THz and (d) 544.5475 THz, corresponding to the resonant frequencies in panels (a) and (b), respectively.

The group-IV emitters are utilized as optical nanoscale thermometers[34–37] as well as a qubit for quantum information processing. In addition to room temperature sensing, a low-power resonantly excited thermometer has been proposed in cryogenic temperatures using a GeV center[38]. Multiple stable PbV centers with nearly identical photon frequencies and transform-limited linewidth developed in this study will lead to micro-scale position dependent thermometers even for the resonant excitation at a specific wavelength, for instance. on observation of vortices in superconducting materials[39–41]. It is also worth mentioning that emitters composed of different group-IV elements have different fluorescence wavelengths, enabling us to perform multi-color thermometry.

## 3. Conclusion

We have obtained the narrow inhomogeneous distributions with standard deviations of several GHz for the PbV centers in diamond by annealing at 2200-2300°C after ion implantation. In contrast, we found that a lower temperature of 1800°C led to a much broader distribution and was insufficient to form PbV centers with a stable PLE spectrum. In the 2200°C annealed sample, the fluorescence from the PbV centers disappeared when irradiating a surrounding spot with 532 nm laser. This is because the PbV centers capture photo-carriers generated under the 532 nm non-resonant laser irradiation, and thus, the PbV centers transition to another charge state. We did not observe this phenomenon for the sample annealed at 2300°C up to an 532 nm irradiation energy of 30 mJ. Finally, we obtained multiple stable PbV centers with nearly identical photon frequencies. This work will promote the research for quantum applications using the PbV center in diamond, and will also contribute to the field of solid-state quantum systems from the viewpoint of the defect control.

## 4. Experimental Section

The PbV centers are formed by implanting Pb ions into IIa-type (001) single-crystal diamond substrates, followed by HPHT annealing[42] using belt-type high-pressure apparatuses[43]. Ion implantation is performed at an acceleration energy of 12 MeV with a fluence of $5 \times 10^8$ cm$^{-2}$. The temperature of the HPHT treatment is set to 1800°C, 2200°C, or 2300°C under a pressure of 7.7 GPa for 20 min. Note that a temperature distribution of approximately ±30°C is expected in the high-pressure cell during the treatment over 2000°C. The temperature variance depends on the sample position in the high-pressure cell and increases as the anneal temperature becomes higher[44]. Furthermore, the temperature variation also occurs when the applied pressure changes, which is caused by the manufacturing precision of high-pressure components, such as the pressure cell, gasket material that seals the pressure, and insulation material. Under a pressure of 7.7 GPa, 2300°C is the limit temperature at which the graphitization can be suppressed and annealing can continue for the duration[45].

The optical properties of the PbV centers are investigated using a home-built CFM system. The samples are placed inside a cryostat and maintained at approximately 6 K throughout all experiments. A 532 nm laser is used for non-resonant excitation, which initializes to the bright negatively charged state of the PbV centers[12]. The fluorescence emitted from the PbV centers under non-resonant excitation is directed either to an avalanche photodiode or a spectrometer. PLE measurements are performed using a tunable dye laser, and the phonon sideband is monitored with a 562 nm long-pass filter. The resonant laser power is 0.7 nW. The non-resonant and resonant lasers are fiber coupled[16,46]. CFM mappings and PLE spectra are recorded using the Qudi Python module[47].

**Acknowledgments**

We thank Y. Chen and Y. Miyamoto for fruitful discussion. This work is supported by JSPS KAKENHI Grant Numbers JP22H04962, the MEXT Quantum Leap Flagship Program (MEXT Q-LEAP) Grant Number JPMXS0118067395, and JST Moonshot R&D Grant Number JPMJMS2062.

# Supporting Information for
# Narrow Inhomogeneous Distribution and Charge State Stabilization of Lead-Vacancy Centers in Diamond

**Ryotaro Abe[1], Peng Wang[1], Takashi Taniguchi[2], Masashi Miyakawa[2], Shinobu Onoda[3], Mutsuko Hatano[1], Takayuki Iwasaki[1,*]**

[1]Department of Electrical and Electronic Engineering, School of Engineering, Institute of Science Tokyo, Meguro, 152-8552 Tokyo, Japan

[2]Research Center for Materials Nanoarchitectonics, National Institute for Materials Science, 305-0044 Tsukuba, Japan

[3]Takasaki Advanced Radiation Research Institute, National Institutes for Quantum Science and Technology, 1233 Watanuki, Takasaki, 370-1292 Gunma, Japan

1. **Spectral diffusion**

During the investigation of the linewidth and inhomogeneous distributions using PLE, some PbV centers in Sample 2200 and Sample 2300 show large spectral diffusion. As shown in Figure S1, a jump with ~400 MHz occurs at each PLE scan for this emitter. Since these kinds of peaks do not form one peak after averaging, the data are not included in the histograms in Figure 2 in the main text.

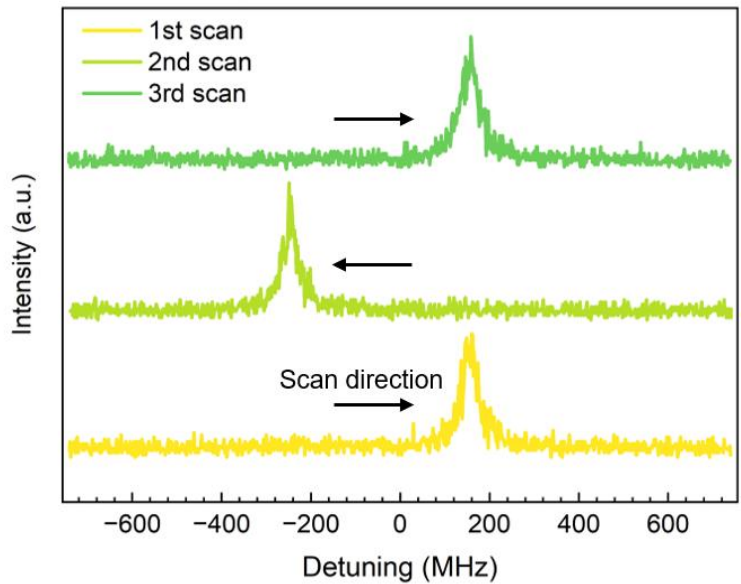

**Figure S1**. Large spectral diffusion.

## 2. Irradiated spots for the photo-carrier generation

For the controlled experiments of the photo-carrier generation in Figure 4 in the main text, the spots ~4.5 μm away from the PbV centers are irradiated by 532 nm laser. Figure S2(a, b) display 532 nm non-resonant CFM images of Sample 2200 and Sample 2300, respectively, showing the target PbV centers and irradiated spots for the photo-carrier generation. In PL spectra at the irradiated spots (Figure S2(c, d)), we only see sharp peaks originating from the first-order Raman scattering from diamond, but not from NV and PbV centers.

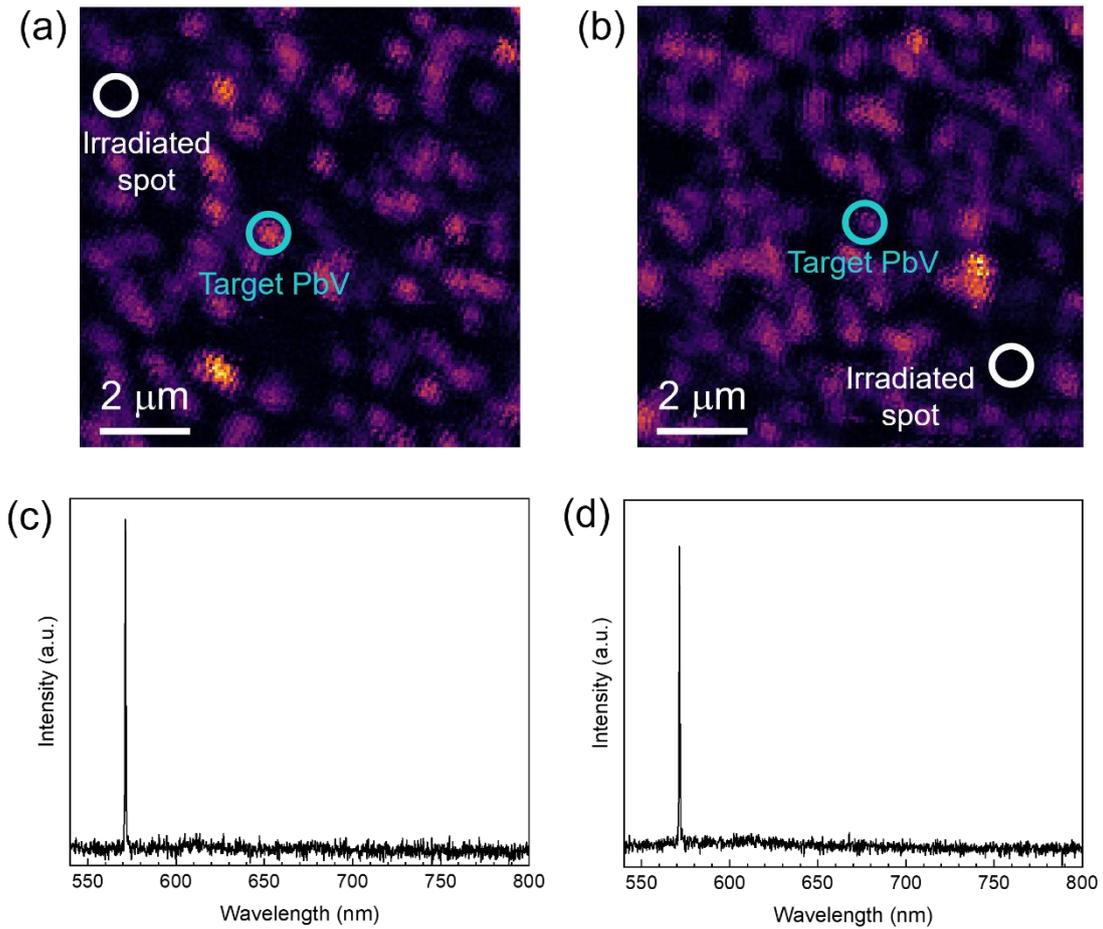

**Figure S2**. Irradiated spots for the photo-carrier generation. CFM images of (a) Sample 2200 and (b) Sample 2300 observed using 532 nm laser. PL spectra from the 532 nm irradiated spots for the photo-carrier generation in (c) Sample 2200 and (d) Sample 2300.

## 3. Investigation of charge state transition by capturing photo-carriers

As shown in the main text, we observe the charge state transition of the PbV center by capturing photo-carriers in Sample 2200. However, we see stable PbV centers in another sample fabricated using the same sample preparation conditions for the fluence and acceleration energy of Pb ion implantation and anneal temperature under high pressure. Figure S3 shows the possibility of observing the bright fluorescence after irradiating a spot away from ~4 μm from a target PbV center using 532 nm laser. Even at an irradiated energy over 2 mJ, the possibility remains 80%, in contrast to Figure 4 in the main text, which shows complete termination at the lower energies.

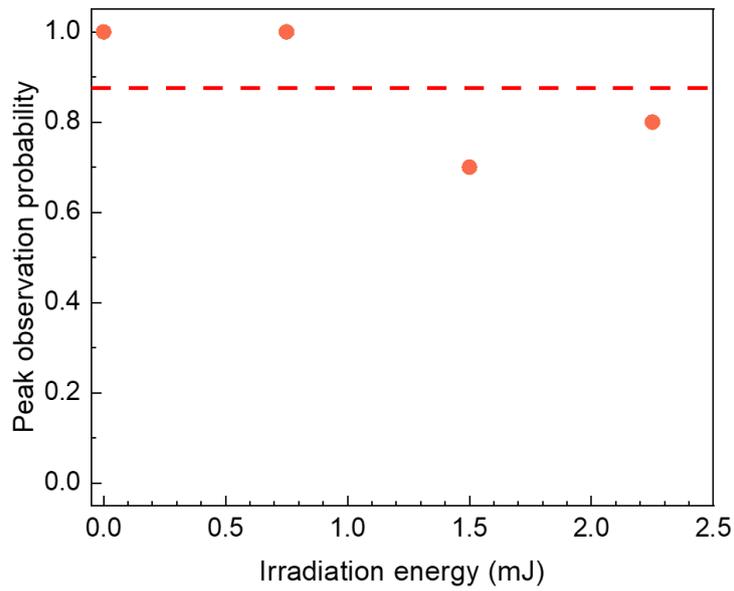

**Figure S3.** Probability of observing the resonant peak as a function of the irradiation energy of the 532 nm laser in another sample fabricated at 2200°C.

## 4. Multiple PbV centers under resonant scan in Sample 2200

In Sample 2200, the PbV centers easily go to a dark state by capturing photo-carriers. However, it is possible to observe multiple PbV centers under resonant scan when using a low power of 532 nm non-resonant laser before resonant scan. Figure S4(a) shows a resonant CFM image obtained after using 30 μW non-resonant laser. We see multiple spots at a specific resonant frequency, indicating that these PbV centers possess similar resonant photon frequencies. In contrast, after using 300 μW non-resonant laser, no emitters are observed under resonant scan (Figure S4(b)).

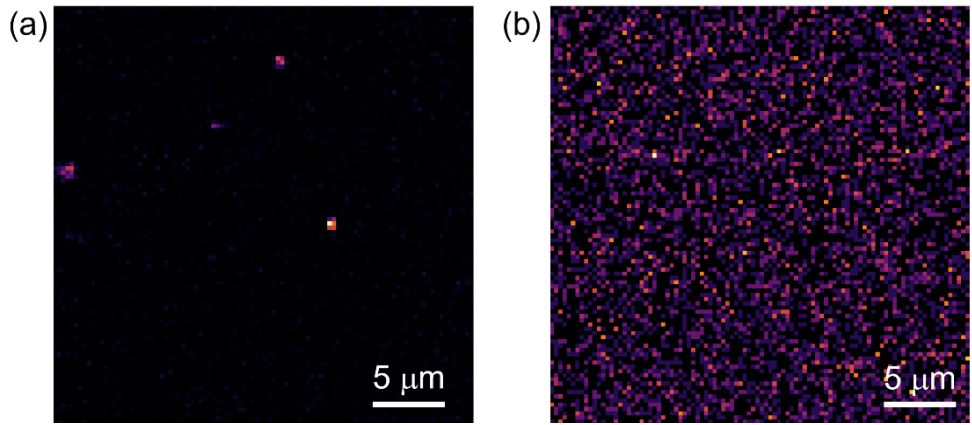

**Figure S4.** Observation of multiple PbV centers under resonant excitation in Sample 2200. Resonant CFM images after irradiated using 532 nm non-resonant laser power of (a) 30 μW and (b) 300 μW.